# Analysis of nonlinear dynamics of a cantilever beam-rigid-body MEMS gyroscope using a continuation method

**Seyed Amir Mousavi Lajimi**
Department of Systems Design Engineering,
Faculty of Engineering, University of Waterloo
200 University Avenue West, Waterloo, Canada, N2L 3G1
samousavilajimi@uwaterloo.ca, samousav@uwaterloo.ca, samirm@sfu.ca

***Abstract -***The nonlinear dynamics of a microbeam-rigid body gyroscope are investigated by using a continuation method. To study the nonlinear dynamics of the system, the Lagrangian of the system is discretized and the reduced-order model is obtained. By using the continuation method, the frequency-response curves are computed and the stability of response is determined.

***Keywords*:** Euler-Bernoulli cantilever beam, eccentric end-rigid-body, microgyroscope, nonlinear dynamics, frequency-response curves, continuation method, bifurcation, AUTO

## 1. Introduction

In vibratory MEMS gyrosocopes the energy is transferred from one direction to its perpendicular direction. Different kinds of designs have been proposed to reduce the fabrication cost and to increase the reliability of MEMS gyroscopes. Esmaili et al. (Esmaeili et al., 2007) modelled a beam-based microgyroscope and studied its fundamental dynamics under linear excitation. Ghommem et al. (Ghommem et al., 2010) added electrostatic nonlinearity to Esmaili et al.'s beam-point mass model. Later Ghommem et al. (Ghommem et al., 2012) studied the nonlinear behavior of the beam-point mass system using the method of multiple scales. Seok and Scarton (Seok and Scarton, 2006) investigated the effect of quality factor on the microbeam gyroscope.

By doing preliminary studies on the effects of adding an end-rigid-body to the cantilever beams, Lajimi et al. (Lajimi et al., 2009) and Lajimi and Heppler (Lajimi and Heppler, 2012b; Lajimi and Heppler, 2012a; Lajimi and Heppler, 2013) have shown that introducing an end-rigid-body with considerable size could be used to modify the dynamics of the structure. Lajimi et al. (Lajimi et al., 2013b; a) have developed the mathematical model of a beam-rigid-body MEMS gyroscope, examined the modal frequencies of the microgyroscope, and studied its linear behavior near primary resonance. In this work, we study the nonlinear behavior of the cantilever beam-rigid body MEMS gyroscope. To this end, we use *XPPAUT* (Ermentrout, 2002), which provides the necessary integration tools and an interface to the continuation and bifurcation package, AUTO (Doedel and Kernevez, 1986; Doedel et al., 1997).

## 2. Reduced-order Model

The Lagrangian of the system includes the kinetic energy of the end-rigid-body and beam, the potential energy of the electrostatic force and beam's elastic deformation, and the nonconservative work of the damping force. The total kinetic energy and potential energy are respectively given by

$$\begin{aligned} K &= K_B + K_M \\ &= \frac{1}{2} M \dot{\boldsymbol{R}} \cdot \dot{\boldsymbol{R}} + \dot{\boldsymbol{R}} \cdot \left( \boldsymbol{\omega}(L,t) \times \int_M \boldsymbol{\rho} \, \mathrm{d}M \right) \\ &\quad + \frac{1}{2} \boldsymbol{\omega}(L,t) \cdot \int_M \boldsymbol{\rho} \times (\boldsymbol{\omega}(L,t) \times \boldsymbol{\rho}) \, \mathrm{d}M + \int_0^L m \, \dot{\boldsymbol{r}}_p \cdot \dot{\boldsymbol{r}}_p \, \mathrm{d}x \end{aligned} \quad (1)$$

and

$$P = -\frac{1}{2}\int_0^L EI\,(v''^2 + w''^2)dx$$
$$+ \frac{1}{2}e\,\epsilon\,h\,(\frac{V_v^2}{g_0 - v(L,t) - e\,v'(L,t)} + \frac{V_w^2}{g_0 - v(L,t) - e\,v'(L,t)}) \quad (2)$$

where $\dot{\mathbf{R}}$ represents the velocity vector of an arbitrary reference point in the end-rigid-body chosen to coincide with the end of the beam and computed relative to the base frame, $\boldsymbol{\omega}(L,t)$ the angular velocity of the beam's section at $L$, $M$ the total mass of the end body, $\boldsymbol{\rho}$ the position vector of an arbitrary point in the end rigid body drawn from the end of the beam to an arbitrary point in the body, $m$ the mass per unit length of the beam, $p$ an arbitrary point on the beam's cross-section, and $\dot{\mathbf{r}}_p$ the velocity vector for the arbitrary point relative to the inertial base frame, see Fig.1, × the vector cross product, · the inner product, $v(x,t)$ and $w(x,t)$ the flexural displacements and $V_v$ and $V_w$ the actuation voltages in the $y$ and $z$ directions, respectively. Initial gap size, the elastic modulus, and the second moment of the area are respectively indicated using $g_0$, $E$, and $I$. The proposed form of the electrostatic potential energy, Eq.(2), is computed at the end-rigid-body's center of mass. Further details are provided in (Lajimi et al., 2013a) and (Lajimi et al., 2013b).

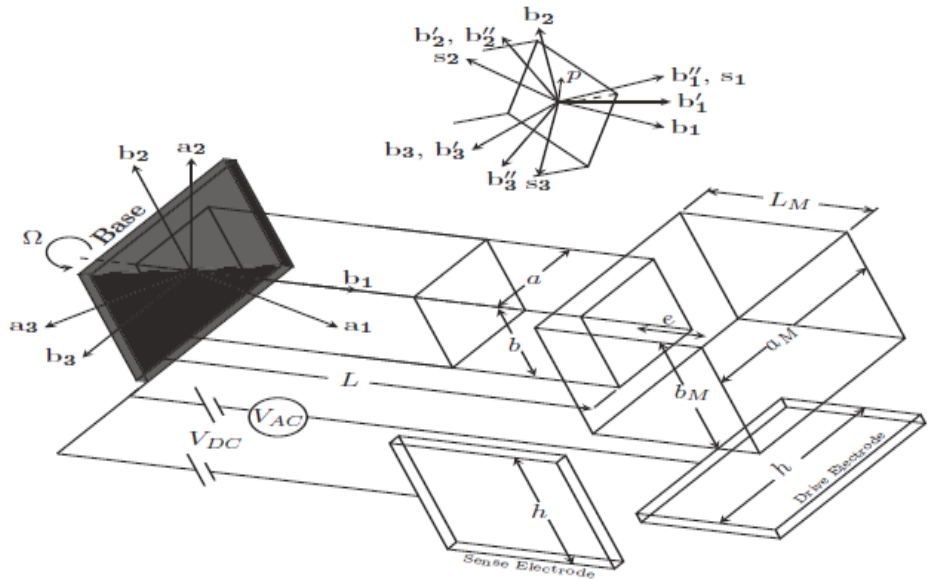

Fig. 1 The microgyroscope: the gyroscope rotates about the longitudinal axis. Systems of inertial $(a_1, a_2, a_3)$, base $(b_1, b_2, b_3)$, and sectional $(s_1, s_2, s_3)$ frames are used to obtain the mathematical model of the MEMS gyroscope. The eccentricity (the distance between the beam's end and the end-rigid-body's center of mass) is denoted by $e$.

Table 1 System parameters

| $L$ ($\mu m$) | $a$ ($\mu m$) | $b$ ($\mu m$) | $L_M$ ($\mu m$) | $a_M$ ($\mu m$) | $b_M$ ($\mu m$) | $h$ ($\mu m$) | $g$ ($\mu m$) | $e$ ($\mu m$) | $E$ (GPa) | $\rho$ ($\frac{kg}{m^3}$) | $\epsilon$ ($\frac{F}{m}$) |
|---|---|---|---|---|---|---|---|---|---|---|---|
| 300 | 5 | 5 | 50 | 20 | 5 | 5 | 2 | 25 | 160 | 2330 | 8.854 × $10^{-12}$ |

To obtain a single-mode approximation of the dynamic response of the system, the response is expressed as

$$w(\hat{\xi}, \hat{t}) = \psi(\hat{\xi})q(\hat{t}) \text{ and } v(\xi, t) = \phi(\hat{\xi})p(\hat{t}) \quad (3)$$

Equation (3) is substituted into the nondimensional form of Lagrangian of the system, $L = K - P$, and Lagrange's equation of motion, see (Meirovitch, 1997) pg.374, is employed to obtain the reduced-order model. The proportional damping terms are added to the equation for each direction. In the sense direction, we compute

$$
\begin{aligned}
&(\Lambda + \alpha\Lambda' + M\phi(1)^2 + 2eM\phi(1)\phi'(1) + J_{33}\phi'(1)^2 + Me^2\phi'(1)^2)\ddot{p}(\hat{t}) + c\Lambda\dot{p}(\hat{t}) \\
&+ \left(\Lambda'' + \Omega^2(\alpha\Lambda' - \Lambda) - M\Omega^2(\phi(1) + e\phi'(1))^2 + (J_{11} \right. \\
&\left. - J_{22})\Omega^2\phi'(1)^2\right) p(\hat{t}) \\
&- \Omega \left(2eM(\psi(1)\phi'(1) + \phi(1)\psi'(1)) + 2e^2 M\Omega\phi'(1)\psi'(1) + 2\Pi \right. \\
&\left. + 2M\phi(1)\psi(1) + (J_{22} + J_{33} - J_{11})\Omega\phi'(1)\psi'(1)\right) \dot{q}(\hat{t}) \\
&= \frac{\epsilon\, e\, \nu\, V_{DC}^2\, \phi(1) + e\phi'(1)}{(1 - \phi(1)p(\hat{t}) - e\phi'(1)p(\hat{t}))^2}
\end{aligned}
\quad (4)
$$

where

$$\Lambda = \int_0^1 \phi(\hat{\xi})^2 d\hat{\xi}, \quad \Lambda' = \int_0^1 \phi'(\hat{\xi})^2 d\hat{\xi}, \quad \Gamma = \int_0^1 \psi(\hat{\xi})^2 d\hat{\xi},$$
$$\Gamma' = \int_0^1 \psi'(\hat{\xi})^2 d\hat{\xi}, \quad \Pi = \int_0^1 \phi(\hat{\xi})\psi(\hat{\xi}) d\hat{\xi} \quad (5)$$

and the nondimensional parameters are defined as

$$\hat{\xi} = \frac{\xi}{L}, \quad \hat{t} = \frac{t}{\kappa}, \quad \hat{e} = \frac{e}{L}, \quad \nu = \frac{6\epsilon h L^4}{E b^4 g^3}, \quad \widehat{M} = \frac{M}{\rho a b L}, \quad \hat{J}_{11} = \frac{J_{11}}{\rho a b L^3},$$
$$\hat{J}_{22} = \frac{J_{22}}{\rho a b L^3}, \quad \hat{J}_{11} = \frac{J_{22}}{\rho a b L^3}, \quad \kappa = \frac{12\rho L^4}{E b^2}, \quad \alpha = \frac{b^2}{12 L^2}, \quad \widehat{\Omega} = \kappa\, \Omega. \quad (6)$$

We obtain a similar equation for the generalized coordinate $q(\hat{t})$ in the drive direction except $V_{DC}$ is replaced with $V_{DC} + V_{AC}$ where $V_{AC} = v_{ac} \cos(\Omega_e t)$ where $\Omega_e$ and $v_{ac}$ indicate the excitation frequency and amplitude, respectively. The system parameters are given in Table 1.

## 4. Numerical results

In Fig.2, frequency-response curves are plotted. To generate the bifurcation diagrams, the system is integrated for a sufficiently large time using *XPPAUT* package. The quality factor is set to 200 and the system is integrated for more than a thousand periods. The maximum and minimum points on the last limit cycle (periodic solutions) are used as inputs to the continuation package. The excitation frequency is then varied to produce frequency-response curves. Both maximum and minimum points on each orbit are plotted in Fig. 2. For small amplitude, $v_{ac} = 0.1$ V, the response is linear, Figs.2 (a) and 2(b). In practice, to reduce irregularities in the performance of gyroscopes, MEMS gyroscopes are preferably operated in the linear range. For large excitation amplitude, $v_{ac} = 1$V, Figs. 2(c) and 2(d), the frequency response curves indicate softening behavior. The response in the sense direction is a combination of a linear branch and a softening branch. The mixed behavior of the sensor should be considered in choosing proper operation range for the microgyroscope.

## 4. Conclusion

In this work, we have presented a reduced-order model of a new cantilever beam-rigid body MEMS gyroscope. By using proper integration-continuation tools, we have studied the system response for two sets of parameters: a small excitation amplitude and a large excitation amplitude. We have shown that for the system under investigation, the response becomes nonlinear for a large excitation amplitude. We have generated the bifurcation diagrams and identified stable and unstable branches of this system. In future works, we study the details of the system response using other methods for a larger range of parameters.

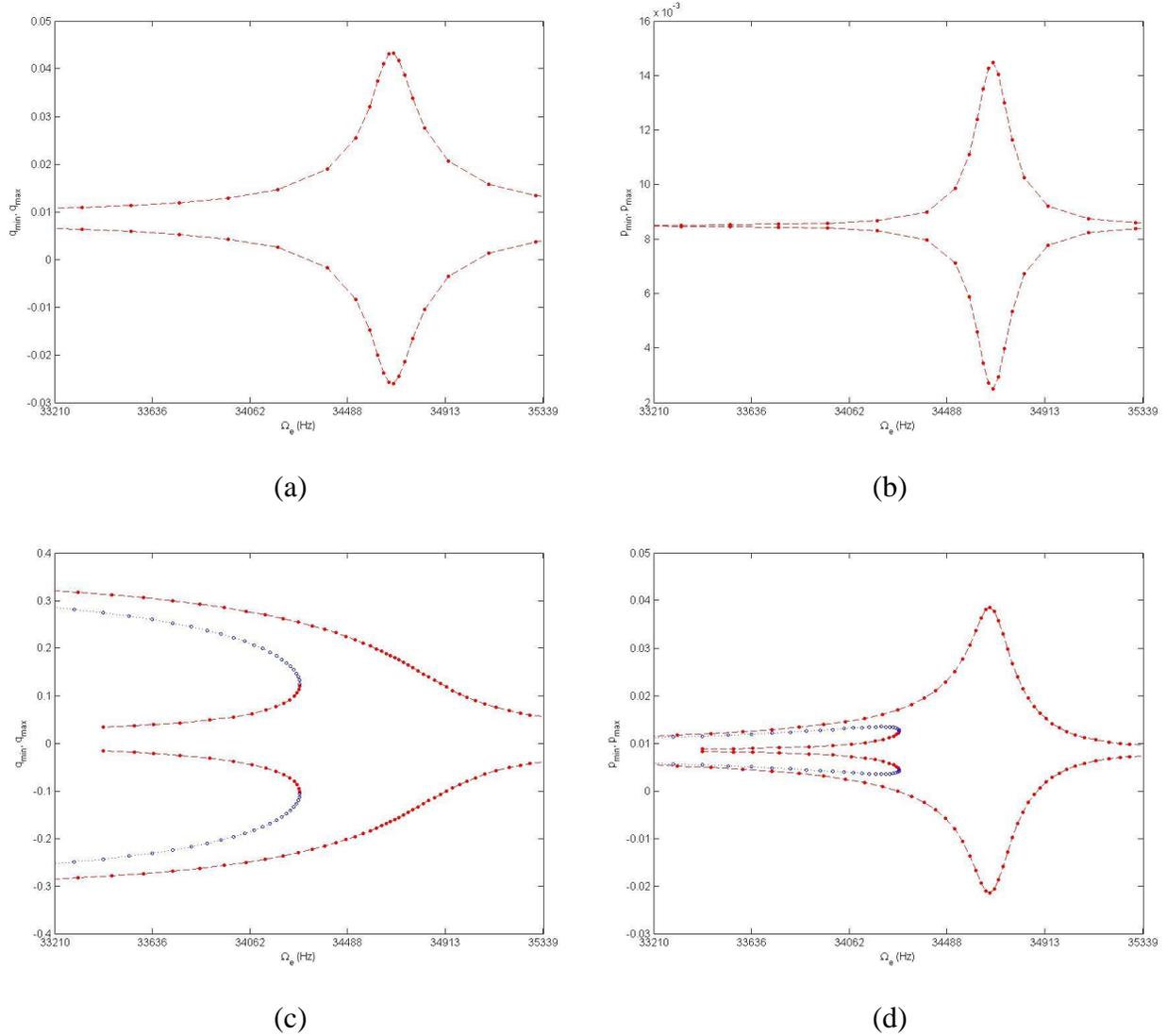

Fig. 2 Bifurcation diagrams plotted for generalized coordinates in the sense, $p(t)$, and drive, $q(t)$, directions. Each plot shows the frequency-response curve when the excitation frequency is varied near the fundamental natural frequency of the system. The excitation amplitude is 0.1 V for (a) and (b) and 1 V for (c) and (d). Solid dots indicate stable branches while empty (blue) circles show unstable branches on (c) and (d). Two points *i.e.* maximum, $q_{max}$, and $p_{max}$, and minimum, $q_{min}$, and $p_{min}$, on each orbit are plotted. The sense and drive amplitudes are denoted by $p$ and $q$, respectively.